\documentclass[reprint,superscriptaddress,nofootinbib,amsmath,amssymb,prl]{revtex4-1}

\usepackage{graphicx}
\usepackage{color}
\usepackage{dcolumn}
\usepackage{bm}
\usepackage{braket}
\usepackage{mathrsfs}
\usepackage{ulem}
\usepackage{multirow}
\usepackage{booktabs}
\usepackage{physics}
\usepackage{amsmath}

\begin{document}
\preprint{APS/123-QED}
\title{Muonium as a probe of point defects in type-Ib diamond}
\author{K.~Yokoyama}
\email{koji.yokoyama@stfc.ac.uk}
\affiliation{
ISIS Neutron and Muon Facility, STFC Rutherford Appleton Laboratory, Didcot, OX11 0QX, United Kingdom
}
\author{J.~S.~Lord}
\affiliation{
ISIS Neutron and Muon Facility, STFC Rutherford Appleton Laboratory, Didcot, OX11 0QX, United Kingdom
}
\author{H.~Abe}
\affiliation{
National Institutes for Quantum Science and Technology (QST), Takasaki, Gunma 370-1292, Japan
}
\author{T.~Ohshima}
\affiliation{
National Institutes for Quantum Science and Technology (QST), Takasaki, Gunma 370-1292, Japan
}
\affiliation{
Department of Materials Science, Tohoku University, Aoba, Sendai, Miyagi 980-8579, Japan
}
\date{\today}

\begin{abstract}
Muonium (Mu), a bound state of a positively charged muon and an electron, can diffuse through crystal lattices and interact with defect centers in insulators and semiconductors.
We demonstrate that this Mu's diffusive property can be used to probe defects in a diamond crystal lattice;
specifically, substitutional nitrogen atoms (N$_\text{s}^0$) and nitrogen-vacancy (NV) centers in type-Ib diamond.
Upon interaction with these defects, Mu can exchange its electron's spin or change its charge state, which result in muon spin relaxation.
However, muons in diamond (and semiconductors in general) can be in a few distinctive muonium states, with each state contributing to the muon signal.
In addition, these states can undergo site and charge exchange interaction, forming a dynamic network.
Hence, to study the Mu interaction with point defects, the muon data have to be deconvoluted to isolate signals from the diffusing species.
To achieve this goal, we have modeled the Mu state exchange dynamics and numerically simulated the time evolution of muon spin polarization by the density matrix method.
With a global curve fit to a set of longitudinal field scan data, one can extract the Mu transition rates that involve interaction with the defect centers.
The diffusing tetrahedral interstitial Mu was found to interact with the paramagnetic N$_\text{s}^0$ center via electron spin exchange.
In contrast, they are converted to form a diamagnetic center upon interaction with the negatively charged NV center.
\end{abstract}
\maketitle
\section{introduction}
Muonium (Mu) is a monatomic bound state made of a positively charged muon $\mu^+$ and a single unpaired electron e$^-$~\cite{Blundell}.
When muons are implanted in materials, they can capture the electron during thermalization.
In semiconductors and insulators, where there is negligible screening, Mu can exist as a metastable state in the timescale of muon's mean lifetime (2.2~$\mu$s)~\cite{Patterson, Chow_Rev, Cox_Rev}.
They are metastable because, in analogy with the highly reactive nature of isolated hydrogen in semiconductors, they should be eventually trapped by lattice defects if they could live long enough.
Among several Mu species known to date, a monatomic Mu state in tetrahedral (T) interstitial sites can be found almost universally in intrinsic tetrahedral semiconductors ($\it e.g.$ diamond, Si, GaAs).
This charge-neutral Mu state, denoted as Mu$_\text{T}^0$, can be characterized with an isotropic hyperfine (HF) coupling constant $A_{\mu}$.
In general, the value of $A_{\mu}$ is in the order of $\sim$GHz and thus indicates that Mu$_\text{T}^0$ is a compact defect center huddling at the center of T-sites.
This light interstitial particle can hop between adjacent equivalent sites and diffuse rapidly through a crystal lattice.
In diamond, thanks to the large interstitial voids, the Mu$_\text{T}^0$ diffusion is considered to be fast with an estimated diffusion coefficient $\approx 10^{-2}$~$\text{cm}^2/\text{s}$ at room temperature (RT)~\cite{Patterson, Holzschuh}.
Hence, Mu$_\text{T}^0$ propagates in a wave-like manner and can interact with defects (if any) in the system.
For instance, as discussed below, there can be spin exchange interaction between Mu$_\text{T}^0$ and paramagnetic centers, or charge exchange interaction between Mu$_\text{T}^0$ and an electron-rich site to form a diamagnetic center.

This paper aims to utilize Mu$_\text{T}^0$ as a potential probe of point defects in semiconductors and demonstrate this principle with type-Ib diamond.
The method is unique because it has sensitivity for detecting not only unpaired electrons but also charged defects without magnetic moments.
Charged defects play a key role in charge carrier recombination processes in semiconductors, which ultimately determine the overall device performance.
Hence, the ubiquity of Mu$_\text{T}^0$ implies its potential application to industrially important semiconductors such as Si and SiC.
In addition, because muons have a well-defined momentum, its implantation depth (and spread) in materials can be readily determined.
Traditional ``surface'' muons used in this study may not give a good spatial resolution.
However, in combination with  a low energy muon source~\cite{Blundell, Bakule}, the method can potentially be a unique probe for defect physics in sub-surface layers.

This paper focuses on Mu interaction with two well-known point defects in type-Ib diamond:
substitutional nitrogen atoms (N$_\text{s}^0$) and nitrogen-vacancy (NV) centers.
The N$_\text{s}^0$ center, known as the ``P1'' centers in diamond studies, substitutes a carbon with nitrogen atom, and thus is in a deep donor state with a single paramagnetic electron~\cite{Loubser}.
Type-Ib diamond, by definition, contains a high concentration of dispersed N$_\text{s}^0$ atoms ($\approx$100~ppm), which are introduced to the host crystal lattice during the growth process.
The NV center, on the other hand, is composed of a nearest neighbor pair of a substitutional nitrogen atom and carbon vacancy site~\cite{Doherty}.
This photoluminescent point defect, its negatively charged species in particular, has attracted immense research interest for decades because of its unique property in hosting a pair of coupled electrons, which can have a total spin of $m_S = \pm$1 or 0.
With its exceptionally long spin coherence time in the order of milliseconds, NV centers in diamond provide a leading platform for various quantum applications such as quantum information processing and high-sensitivity local field probing~\cite{Barry}.

For NV center generation, it is now a standard procedure to start with type-Ib diamond as a host substrate, which is then exposed to an electron beam for generating carbon vacancy (V) sites~\cite{Ishii}.
Subsequent annealing activates the V site, which then diffuses and pairs with a N$_\text{s}^0$ atom to form a NV center.
Generated NV centers are in a neutral NV$^0$ state, which requires one more electron to form the electron pair and become a negatively charged NV$^-$ state.
The key property with NV centers is that, because the energy level of NV$^-$ is lower than that of NV$^0$, the NV$^0$ state can accept an electron to form a NV$^-$ state~\cite{Larsson, Lofgren}.
It is widely known that this additional electron is provided by a charge-transfer from a nearby N$_\text{s}^0$ center~\cite{Lofgren}.
Obviously, generating a high concentration of stable NV$^-$ states is important for the quantum applications;
to date, the concentration can reach as high as 10$^{18}$~cm$^{-3}$ with NV centers fully converted to the negative state~\cite{Shinei}.
We consider that this system provides an ideal test bench for studying Mu interacting with defects in diamond.

In diamond, two neutral Mu centers have been identified as metastable states: Mu$_\text{T}^0$ and Mu$_\text{BC}^0$.
The latter is a neutral Mu state localized at the center of a C-C bond.
The Mu$_\text{BC}^0$ center is therefore axially symmetric about $\langle$111$\rangle$, with $A_{\parallel}$ and $A_{\perp}$ representing HF components parallel and perpendicular to the $\langle$111$\rangle$ axis respectively.
In diamond (similarly with other tetrahedral semiconductors), these HF constants are an order of magnitude smaller than $A_{\mu}$
$\it i.e.$ $A_{\mu}$ = 3711~MHz, $A_{\parallel}$ = 168~MHz, and $A_{\perp}$ = $-393$~MHz~\cite{Odermatt}.
Because of the electron orbital spread along the bond, Mu$_\text{BC}^0$ stretches the C-C bond and hence is polaronic.
As a result of this self-trapping, Mu$_\text{BC}^0$ in diamond is immobile up to 1000~K~\cite{Madhuku}.
Reportedly, Mu$_\text{T}^0$ and Mu$_\text{BC}^0$ exist in diamond with $f(\text{Mu}_\text{T}^0)$ = 0.7 and $f(\text{Mu}_\text{BC}^0)$ = 0.2, where $f$ denotes a promptly formed fraction of a Mu state~\cite{Connell};
the rest, $1-f(\text{Mu}_\text{T}^0)-f(\text{Mu}_\text{BC}^0)$, is considered to be in diamagnetic states.

\begin{figure}
\includegraphics{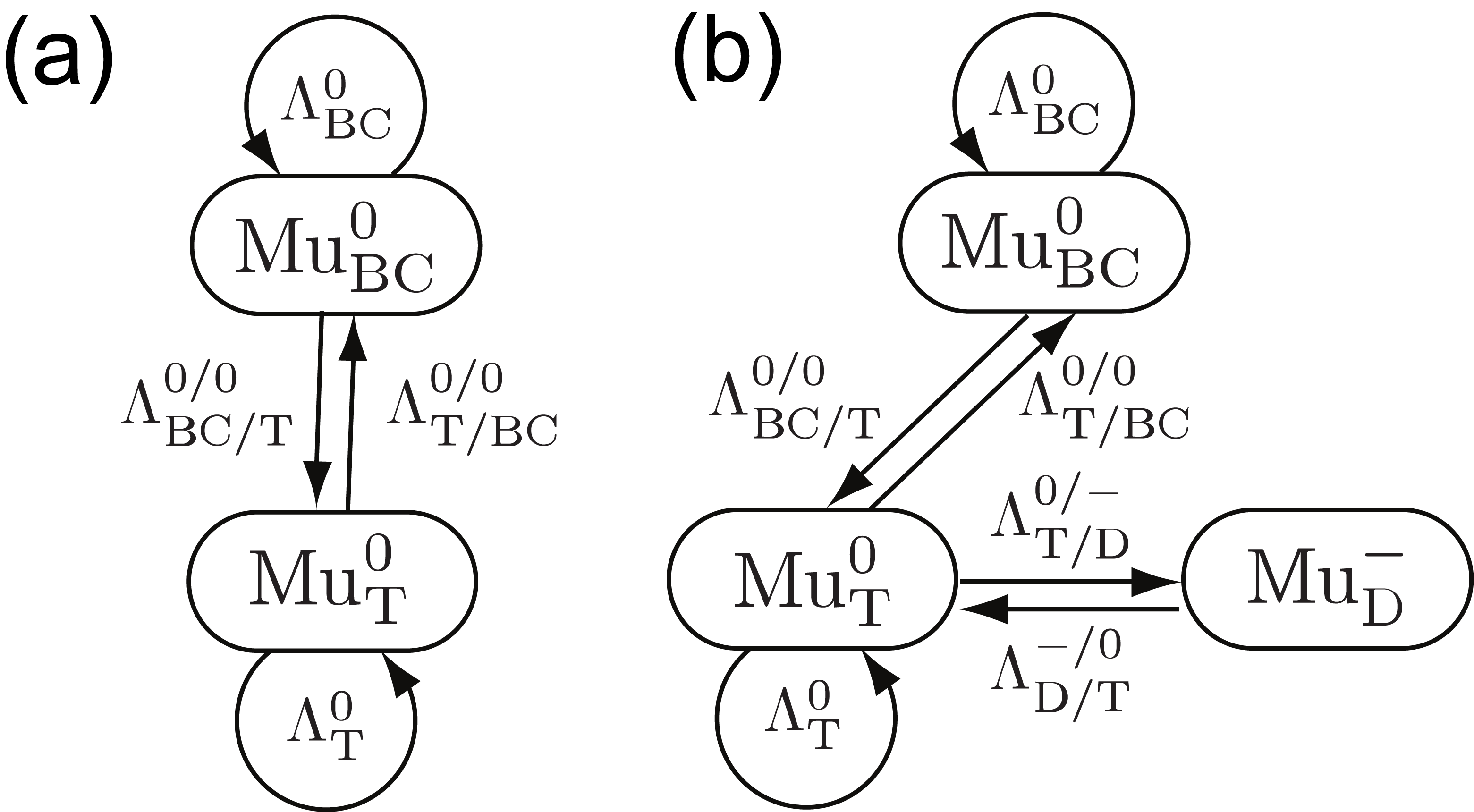}
\caption{\label{fig:Fig_Mu_diagrams}
Model of Mu state exchange in the (a) Pristine and (b) NV sample.
A superscript and subscript of $\Lambda$ indicate the charge state and Mu site respectively, with a forward slash between before and after a transition.
}
\end{figure}

The muon spin relaxation ($\mu$SR) time spectrum is therefore a convolution of signals from these Mu centers, which can spontaneously exchange their state.
In addition, their muon spin can relax upon Mu interacting with the host system.
Hence, in order to fully understand underlying Mu exchange processes and characterize interaction with the host crystal lattice, it is imperative to numerically simulate the $\mu$SR time spectra.
Below, we identify a model of the dynamically exchanging Mu system in diamond and characterize Mu interaction with the two point defects.
With this model, the density matrix method can calculate the time evolution of the muon spin~\cite{Lord1, Lord2}, which is used to curve fit $\mu$SR time spectra from the experiment.
\footnote{
Historically, a model of Mu dynamics has been developed to study $\mu$SR signals from Si
(similar to the models shown in Fig.~\ref{fig:Fig_Mu_diagrams})~\cite{Chow_Rev, Cox_Rev}.
Modern computing resources have enabled calculations with these models.
Though there are not many studies done yet, Fan {\it et al.}, for example, calculated the muon spin evolution in illuminated Si with a slightly different method but based on the same principle as this paper~\cite{Fan}.
}
Zeeman splitting by a longitudinal field (LF) application decouples Mu's HF interaction and helps differentiating the Mu states.
The interaction with the host crystal lattice can be extracted from relaxation rates of Mu spins.

\begin{figure}
\includegraphics{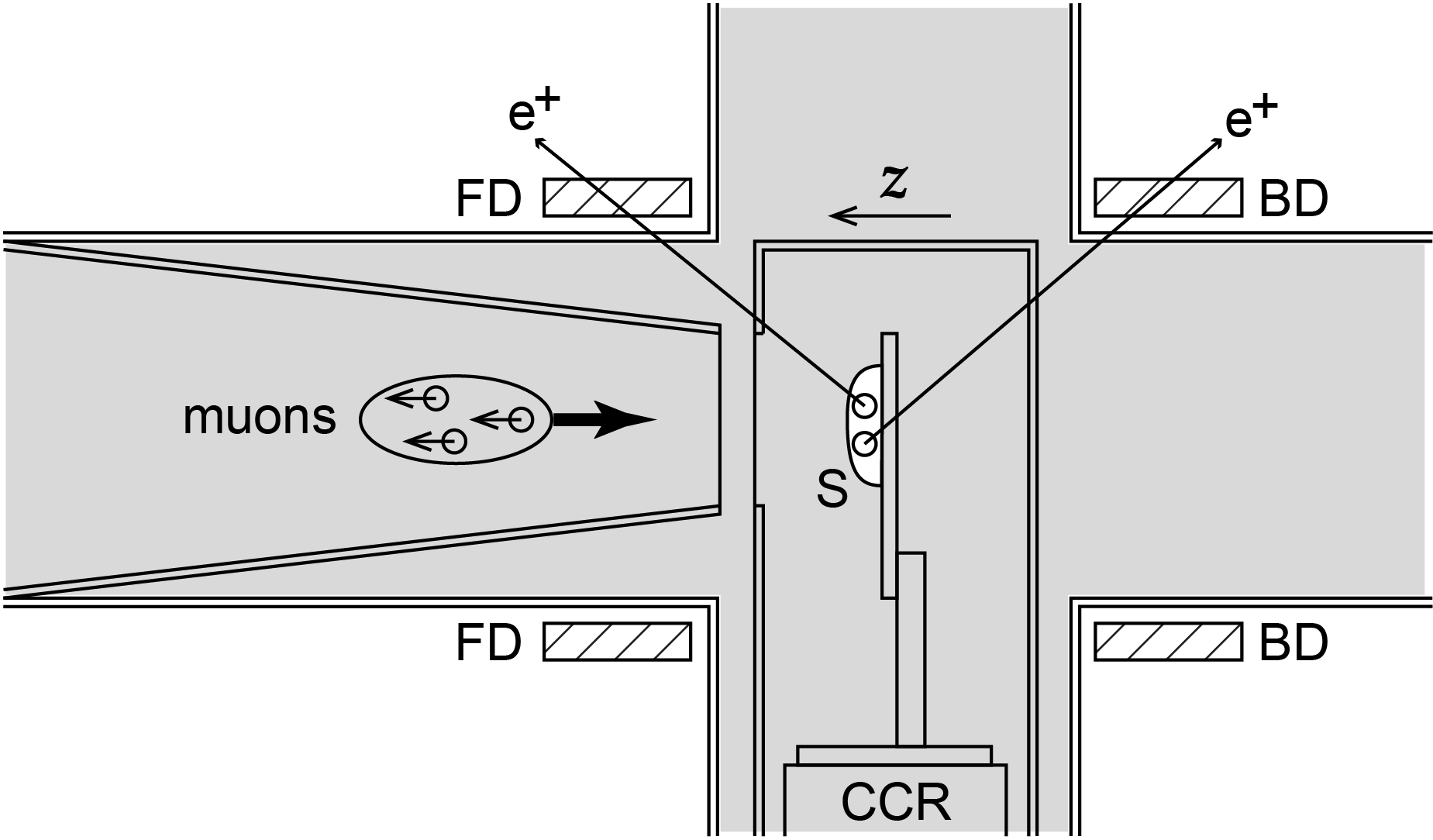}
\caption{\label{fig:Fig_schematic}
Schematic diagram of the experimental setup.
Muons propagate through the beam snout and irradiate a sample (``S'').
Muon spins are fully polarized in a direction antiparallel to their momentum vectors (shown by the arrows).
The sample was mounted on a closed cycle refrigerator (``CCR'') with a radiation shield.
Forward (``FD'') and backward (``BD'') detector segments surrounding the vacuum chamber (shown in gray) detect high-energy decay positrons.
There are a total of 96 detectors in EMU with a count rate of 120 million events per hour.
A Helmholtz coil (not shown here) applies a uniform longitudinal field at the sample position along the $\vb*{z}$ axis.
}
\end{figure}

\begin{figure*}
\includegraphics{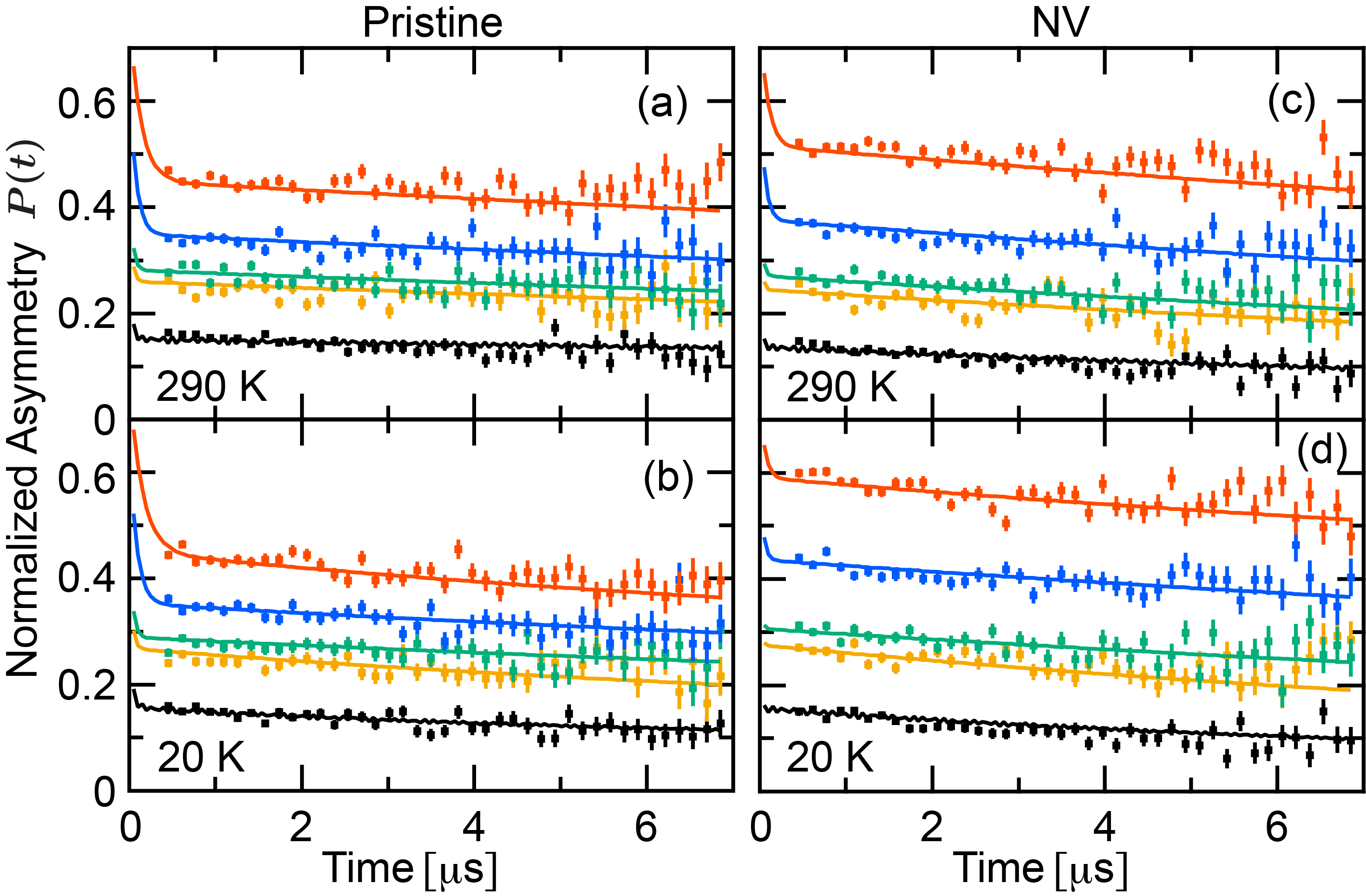}
\caption{\label{fig:Fig_TS}
$\mu$SR time spectra measured on the Pristine [(a) and (b)] and NV [(c) and (d)] sample at 290 and 20~K under five representative LF fields {\it i.e.}
0 (black), 200 (yellow), 600 (green), 2000 (blue), and 4000~G (red).
Five million muon decay events were averaged for each spectrum.
Error bars are due to the random decay of the muons, mean lifetime 2.2~$\mu$s, and the Poisson counting statistics of the decay positrons (hence the shorter error bars in the earlier time bins).
The full muon asymmetry in this EMU setup, $\approx$22.0~\%, was used to normalize the spectra and get $P(t)$ 
({\it i.e.} a full asymmetry gives a unity on the y-axes).
Solid lines denote results of the curve fitting (see text).
The numbers of fit parameters are six for (a) and (b), and five for (c) and (d) (see TABLE~\ref{table:fit_results}).
Reduced chi-squared values were (a) 1.196, (b) 1.174, (c) 1.483, and (d) 1.279.
In high LF, there can be systematic errors associated with decay positrons spiralling around field lines,
resulting in a small shift in the offset.
}
\end{figure*}

\section{experiments and results}
We measured two single-crystal diamond samples in granular form with an average grain diameter of 1.2~mm.
One is as-received, synthetically grown type-Ib diamond containing a N$_\text{s}^0$ concentration of approximately 100~ppm, which has been estimated from electron paramagnetic resonance measurements~\cite{Smith, Cox_A}.
For simplicity, we call this sample ``Pristine''.
The other is diamond with NV centers (hereafter called ``NV'').
For making the NV sample, the Pristine sample was exposed to an electron beam for the vacancy site generation and subsequently heat-treated at 900~${}^\circ$C for 2~hours for NV center generation.
The resulting NV concentration is estimated to be $\approx$4$\times$10$^{17}$~cm$^{-3}$ (or 2.5~ppm).
The conversion efficiency from a NV$^0$ to NV$^-$ state depends on the availability of N$_\text{s}^0$ atoms for each NV center~\cite{Lofgren}.
In our sample, $>$90~\% of NV centers are considered to be in the NV$^-$ state at RT~\cite{Shinei}.
Each sample weighed about two grams.

The experiment was carried out using the high-rate, general-purpose $\mu$SR spectrometer, EMU~\cite{Giblin}, in the ISIS Pulsed Neutron and Muon Source located at the STFC Rutherford Appleton Laboratory in the UK~\cite{DataDoi}.
As shown in Fig.~\ref{fig:Fig_schematic}, the experiment was carried out in a typical LF-$\mu$SR geometry, where the magnetic field is applied parallel to the direction of the initial muon spin polarization~\cite{Blundell}.
Measurements were taken for fields ranging between 0 and 4000~Gauss.
The sample was loaded in a packet made of a piece of thin silver (Ag) foil, which was then mounted on a Ag plate with a small amount of grease for better thermal conductivity.
The plate was then mounted on a closed cycle refrigerator for cooling.
A 100~\% spin polarized muon beam with a kinetic energy of 4~MeV (known as ``surface'' muons) was incident on the sample.
The muon range ({\it i.e.} an areal density to fully stop muons) in this setup is $\approx$200~mg/cm$^2$, and the beam diameter is $\approx$1~cm FWHM.
Therefore, the 2-gram samples in the square Ag packet, whose area was 2.5$\times$2.5~cm$^2$, should be enough to stop most of incoming muons in the sample volume.

\begin{figure*}
\includegraphics{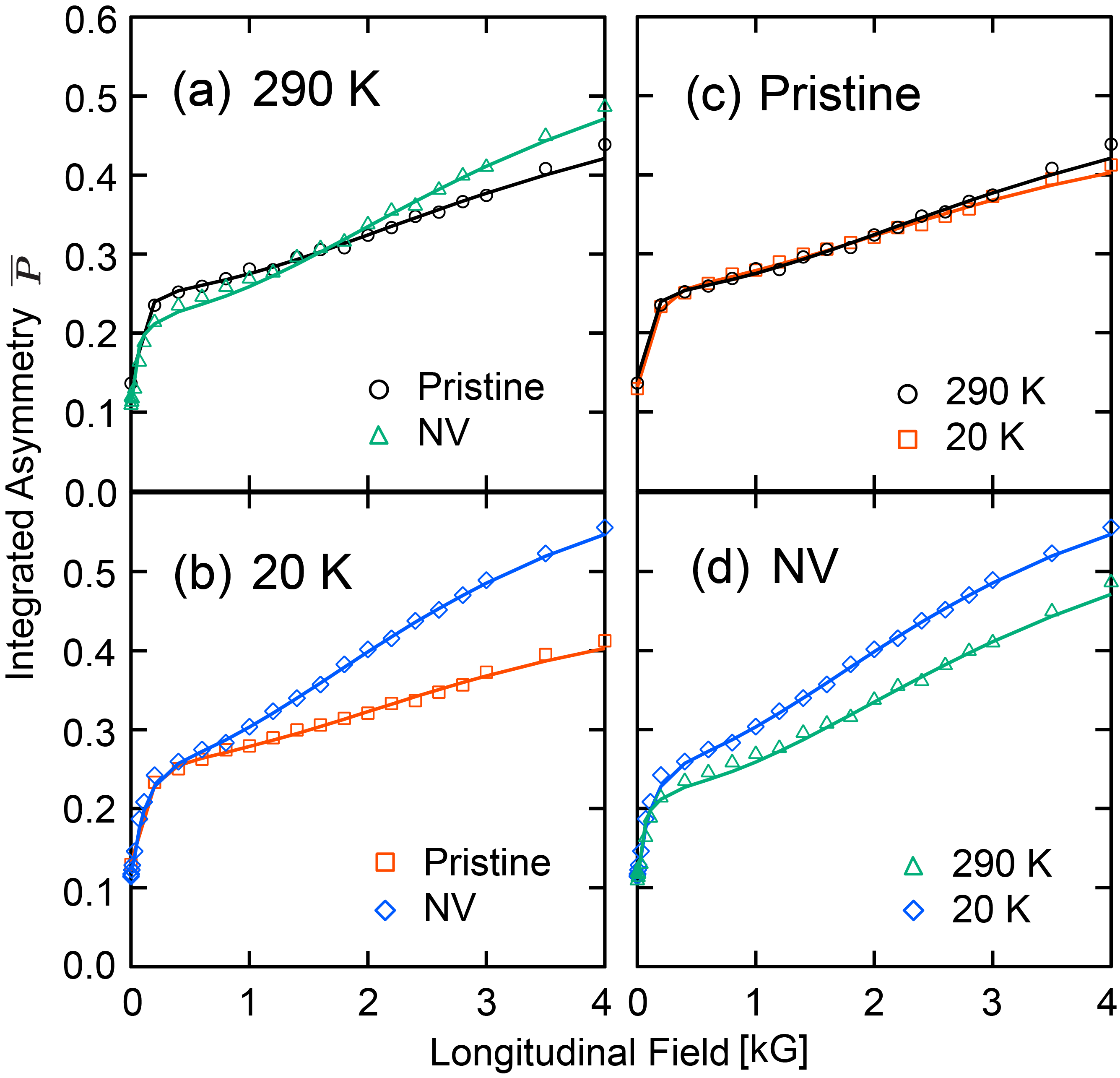}
\caption{\label{fig:Fig_int}
Repolarization curves ($\overline{P}$ {\it vs.} LF) for the Pristine and NV sample measured at (a) 290 and (b) 20~K.
Solid lines denote the curve fit results (see text).
The same set of data are replotted for the (c) Pristine and (d) NV sample for comparing their temperature dependence.
For clarity, the plots are color-coded such that:
black circles for Pristine at 290~K,
red squares for Pristine at 20~K,
green triangles for NV at 290~K,
and blue diamonds for NV at 20~K.
Note that the difference in the offsets ($\approx$0.02) in Fig.~(a) and (b) are associated with background signals from the silver materials.
Even if the two samples were mounted in the same way, there always is a small difference in muon beam coverage.
The Supplemental Material has Fig.~(a) and (b) after the offset correction~\cite{SM}.
With this adjustment, the decoupling of Mu$_\text{T}^0$ can be more clearly compared between Pristine and NV.
}
\end{figure*}

Figure~\ref{fig:Fig_TS} shows representative $\mu$SR time spectra measured on the Pristine and NV sample at 290 and 20~K.
Zeeman energy applied by the longitudinal field decouples HF interactions, resulting in the recovery of muon spin polarization.
To measure this decoupling behavior, the time spectra have been integrated and averaged, and plotted as a function of LF.
Namely, the quantity,
$\overline{P} = \lbrack \sum\nolimits_{i=1}^{M} P(t_i) \rbrack /M$,
has been calculated for each field, where, in the numerator, the normalized muon spin asymmetry $P$ at time $t_i$ are summed over all bins from $i$ = 1 to $M$.
\footnote{
This simple sum of polarizations for $\overline{P}$, however, is applicable only for simulations.
For experimental data, in order to take account of the larger error bars in later time bins (see Fig~\ref{fig:Fig_TS}), the following quantity is calculated instead:
\begin{gather}
\notag
\overline{P} = \frac{N_F - \alpha N_B}{N_F + \alpha N_B}\\
\notag
N_F = \sum_{i=1}^{M} n_F(t_i); N_B = \sum_{i=1}^{M} n_B(t_i),
\end{gather}
where $\alpha$ is a geometrical correction factor for detectors~\cite{Blundell}, and $n_F(t_i)$ and $n_B(t_i)$ are collective positron counts in forward and backward detectors at $t_i$ respectively. 
}
These so-called ``repolarization curves'' are shown in Fig.~\ref{fig:Fig_int}.
Since most of the muon spin polarization is lost at 0~G ($\overline{P} \sim 0.1$), it is clear that there are fast dynamics in the system, which depolarizes Mu in their triplet state (with its total spin $m_F$ = +1).
This fast depolarization appears to be too fast to observe at pulsed sources.
The signals are then quickly recovered by the field to 400~G, followed by the slower repolarization that continues to 4000~G.
Based on the previous studies~\cite{Cox_Rev, Connell}, Mu$_\text{BC}^0$ and Mu$_\text{T}^0$ can be associated with the repolarization in the low and high field respectively.
Clearly, the repolarization behavior in the high-field range is dependent on the sample and temperature, presumably reflecting changes in how Mu$_\text{T}^0$ interacts with the host systems.
However, in order to fully understand physics behind the curves, we need to simulate the $\mu$SR time spectra with a model.

Note that the applied field should hardly change the spin state distribution of NV$^-$ centers, and hence its interaction with Mu.
Since this experiment was carried out without optical excitation, all NV$^-$ centers were in their ground state and not spin-polarized.
The ground state is split by 2.87~GHz between the $m_S = 0$ and degenerate $\pm1$ states~\cite{Doherty}.
An external magnetic field $B_0$ along the defect axis splits the $m_S = \pm1$ levels by $2 \gamma_{NV} B_0$, 
where $\gamma_{NV}$ = 28~GHz/T is the gyromagnetic ratio of a NV$^-$ center.
The spin states are populated according to the Boltzmann distribution.
Both the zero-field and Zeeman splitting in 4000~G are two orders of magnitude smaller than the thermal energy at 20~K.
Therefore, the field application (and potential magnetic impurities) exerts a negligible impact on the population of the sublevels.

\section{density matrix simulations}
To build the model, we simply assumed that
Mu$_\text{T}^0$ and Mu$_\text{BC}^0$ are the only active Mu states in the system, and
the rest is diamagnetic giving a small constant background.
In diamond, Mu interaction with nearby nuclear spins is negligibly small because there is no nuclear spin in $^{12}$C nuclei, which take up $\sim$99~\% of naturally occurring isotopes.
Hence, the system's interaction with the bath takes place only via the electron spin exchange (hence inducing relaxation) and Mu state transitions.
This model is shown in Fig.~\ref{fig:Fig_Mu_diagrams}(a).
Here, there are site exchange interactions, where Mu$_\text{T}^0$ can be spontaneously converted to Mu$_\text{BC}^0$ and vice versa.
Each transition process can be characterized by an exchange rate $\Lambda$.
In addition, Mu$_\text{T}^0$ and Mu$_\text{BC}^0$ can exchange their electron's spin with the bath at rates, $\Lambda_\text{T}^0$ and $\Lambda_\text{BC}^0$, respectively.
The diamagnetic component is part of the simulation but assumed only to give a time-independent background.
Note that the diffusive motion of Mu$_{\text{T}}^0$ is not part of the model;
however, these rates can carry contribution from the Mu motion and interaction with the host crystal lattice.

The simulation was carried out using QUANTUM, a Python program to solve the time evolution of the muon spin using the density matrix method.
As full details of the program are available elsewhere~\cite{Lord1, Lord2}, here, we summarize its computational flow.
The program solves an equation of motion,
\begin{equation}\label{eqn_motion}
\frac{\partial \rho}{\partial t} = \lbrack \mathcal{H}, \rho \rbrack,
\end{equation}
where $\rho$ and $\mathcal{H}$ are a density matrix operator and Hamiltonian respectively.
A physical quantity at time t, such as the muon spin polarization $P(t)$, can be calculated from
\begin{equation}\label{eqn_trace}
\langle P(t) \rangle = \text{Tr} \lbrack \rho(t) P \rbrack.
\end{equation}
The program applies a basis set and expands Eq.~(\ref{eqn_motion}) with the completeness relation,
$\sum_{k}^{N} \ket{k} \bra{k} = 1$,
such that
\begin{equation}\label{eqn_motion_expanded}
\frac{\partial \rho_{ij}}{\partial t} = \sum_{k}^{N} (\mathcal{H}_{ik} \rho_{kj} - \rho_{ik} \mathcal{H}_{kj}),
\end{equation}
where $\rho_{ij} = \bra{i} \rho \ket{j}$, $\mathcal{H}_{ij} = \bra{i} \mathcal{H} \ket{j}$, and $N$ is the dimension of the basis.
Here, the basis set has each spin as a good quantum number.
For a Mu system, for example, the basis states are
$\ket{1} = \ket{++}$, $\ket{2} = \ket{+-}$, $\ket{3} = \ket{-+}$, and $\ket{4} = \ket{--}$ ($N = 4$),
where the first $+$ or $-$ sign refers to the muon and the second to the electron.
The code then flattens the matrix $\rho$ into a vector $\vb*{R}$
(there are $N \times N =$ 16 vector components in this example).
Since the right hand side of Eq.~(\ref{eqn_motion_expanded}) is a linear combination of $\rho_{ij}$, Eq.~(\ref{eqn_motion_expanded}) can be written as
\begin{equation}\label{eqn_motion_flatten}
\frac{\partial \vb*{R}}{\partial t} = X \vb*{R},
\end{equation}
where $X$ is a large square matrix with $\mathcal{H}_{ij}$ in its components (this is a $16 \times 16$ matrix for the example system).
Clearly, Eq.~(\ref{eqn_motion_flatten}) is a matrix expression of coupled linear ordinary differential equations, and readily solvable as an eigenvalue problem.
By diagonalizing $X$, Eq.~(\ref{eqn_motion_flatten}) can be written as
\begin{equation}\label{eqn_motion_diagonalized}
\frac{\partial \tilde{\vb*{R}}}{\partial t} = D \tilde{\vb*{R}},
\end{equation}
where $D$ is a diagonal matrix with the eigenvalues in its diagonal components, and $\tilde{\vb*{R}} $ is a projection of $\vb*{R}$ to the new coordinates.
The solution is then of the form
\begin{equation}\label{eqn_motion_solution}
\tilde{\vb*{R}}(t) = e^{Dt} \tilde{\vb*{R}}(0).
\end{equation}

For a system with multiple Mu states, the code appends the vectors $\vb*{R}$ for each state (as $\rho_{T}$ for Mu$_{\text{T}}^0$ and $\rho_{BC}$ for Mu$_{\text{BC}}^0$) to give a single long vector.
Accordingly, for $X$, it joins the matrix for each Mu state as block diagonals and gives a single large matrix, while setting the off-diagonal blocks to 0.
For our model in Fig.~\ref{fig:Fig_Mu_diagrams}(a), but before considering any Mu state transitions and spin relaxations,
$X$ will be then of the form
\begin{equation}\label{eqn_matrix_X_no_transitions}
X=
\left[
\begin{array}{c|c}
   X_T & 0 \\ \hline
   0 & X_{BC}
\end{array}
\right],
\end{equation}
where $X_T$ and $X_{BC}$ are the $X$ matrices for Mu$_{\text{T}}^0$  and Mu$_{\text{BC}}^0$ respectively.
In this context, the promptly formed Mu fractions discussed above can be written as
$f(\text{Mu}_\text{T}^0) = \text{Tr} \lbrack \rho_{T}(t = 0)\rbrack$ and
$f(\text{Mu}_\text{BC}^0) = \text{Tr} \lbrack \rho_{BC}(t = 0)\rbrack$.

The Hamiltonian for Mu$_{\text{T}}^0$ is given by
\begin{equation}\label{eqn_Hamiltonian_T}
\frac{\mathcal{H}_T}{\hbar} = -\gamma_e \vb*{S} \cdot \vb*{B} -\gamma_\mu \vb*{I} \cdot \vb*{B} + A_\mu \vb*{I} \cdot \vb*{S},
\end{equation}
where
$\hbar$ is the reduced Planck constant, and
$\gamma_e$ and $\gamma_\mu$ are the electron and muon gyromagnetic ratio respectively.
Here, we have also defined the spin operators, $\vb*{S}$ and $\vb*{I}$, for the electron and muon respectively, and the applied magnetic field $\vb*{B}$.
In the current case, $\vb*{B}$ is parallel to the initial muon spin polarization and directed along $\vb*{z}$ (see Fig.~\ref{fig:Fig_schematic}).
The first two terms in Eq.~(\ref{eqn_Hamiltonian_T}) account for the Zeeman energy,
and the last term represents the isotropic hyperfine interaction in Mu$_{\text{T}}^0$.
For Mu$_{\text{BC}}^0$, the Hamiltonian is given by
\begin{equation}\label{eqn_Hamiltonian_BC}
\frac{\mathcal{H}_{BC}}{\hbar} = -\gamma_e \vb*{S} \cdot \vb*{B} -\gamma_\mu \vb*{I} \cdot \vb*{B}
+ A_{\parallel}S_{z^{\prime}}I_{z^{\prime}} + A_{\perp}(S_{x^{\prime}}I_{x^{\prime}} + S_{y^{\prime}}I_{y^{\prime}}).
\end{equation}
The last three terms describe the axially symmetric hyperfine interaction with the symmetry axis $\vb*{z}^{\prime}$,
which can be tilted at an angle from $\vb*{z}$.
Indeed, the program calculates Mu$_{\text{BC}}^0$ along the four bond axes separately
({\it i.e.} $\langle 111 \rangle$,
$\langle 1 \overline{1} \overline{1} \rangle$,
$\langle \overline{1} 1 \overline{1} \rangle$, and 
$\langle \overline{1} \overline{1} 1 \rangle$).
Hence, to be exact, there should be four block diagonals for Mu$_{\text{BC}}^0$ in Eq.~(\ref{eqn_matrix_X_no_transitions}).
In addition, to account for the granular sample form, the code uniformly distributes the orientation over the sphere.
Finally, in the actual computation, $A_{\parallel}$ and $A_{\perp}$ for Mu$_{\text{BC}}^0$ are decomposed into
an isotropic ($A_{\text{iso}}$) and traceless dipolar ($A_{\text{dip}}$) term given by
$A_{\text{iso}} = (A_{\parallel} + 2 A_{\perp})/3$ and 
$A_{\text{dip}} = 2(A_{\parallel} - A_{\perp})/3$.

The code applies a relaxation of the electron spin in Mu$_{\text{T}}^0$ by adding a relaxation rate $\lambda$ to some of the elements in $X_T$.
Here, we discuss a system made only of Mu$_{\text{T}}^0$ as an example,
but the same discussion can be applied to Mu$_{\text{BC}}^0$.
Because Mu$_{\text{T}}^0$ interacts with unpolarized electrons in the system (hence, $\lambda = \Lambda_{T}^0/2$), 
we expect $\rho (t \to \infty) = \frac{1}{N} \vb{1}$ ({\it i.e.} it is eventually completely unpolarized),
while maintaining the condition, $\text{Tr} \lbrack \rho(t) \rbrack = 1$.
Therefore, the program adds $-\lambda$ to the whole diagonal of $X_T$ and then $+\lambda/2$ to selected elements.
The first operation gives the relaxation of $\rho_{ij}$ by adding $-\lambda \rho_{ij}$ in the differential equation,
whereas the second operation keeps the balancing of $\rho$'s and maintain the normalization condition.
Understanding this procedure is easier by considering an example with a single spin system,
which is available in the Supplemental Material for interested readers~\cite{SM} .

The same idea can be applied to add the Mu state transition to the $X$ matrix.
For diagonal elements of a Mu state, the program subtracts the sum of of transition rates out of the state.
In contrast, to account for the incoming rate to the other Mu state, the code adds it to the diagonal elements in the off-diagonal blocks joining these states.
For example, if we consider the transition from Mu$_{\text{T}}^0$ to Mu$_{\text{BC}}^0$, which is characterized by $\Lambda_\text{T/BC}^{0/0}$, Eq.~(\ref{eqn_matrix_X_no_transitions}) is modified such that
\begin{equation}\label{eqn_matrix_X_state_transition}
X=
\left[
\begin{array}{c|c}
   X^{T}_{ii} - \Lambda_\text{T/BC}^{0/0} & 0 \\ \hline
   0_{ii} + \Lambda_\text{T/BC}^{0/0} & X_{BC}
\end{array}
\right],
\end{equation}
where $X^{T}_{ii}$ and $0_{ii}$ represent diagonal elements in the $X_{T}$ and $0$ matrix respectively
($X^{T}_{ij}$ and $0_{ij}$ for $i \neq j$ remain the same).
Again, Eq.~(\ref{eqn_matrix_X_state_transition}) is only for illustration purposes because the actual $X$ matrix should contain the four block diagonals for Mu$_{\text{BC}}^0$.

After including all Mu states and dynamics in $X$, the program solves Eq.~(\ref{eqn_motion_flatten}) as an eigenvalue problem with a numerical library subroutine.
In general, $X$ matrices are not Hermitian, and obtained eigenvalues are complex.
Hence, the solution is obtained in the form
\begin{equation}\label{eqn_solution}
\tilde{\vb*{R}}(t) = \sum_{k} e^{-\lambda_k + i \omega_k}R_k(0),
\end{equation}
where the real ($\lambda_k$) and imaginary ($\omega_k$) part of the eigenvalues represent a damped oscillation of the vector component $R_k$.
Finally, Eq.~(\ref{eqn_trace}) is evaluated for an observable, in our case, for $P(t)$.
The program returns a normalized time domain signal, which can be used to curve fit a measured $\mu$SR spectrum.
Global curve fitting on {\it all} of $\mu$SR time spectra in a LF data set enables us to distinguish contributions from Mu$_{\text{T}}^0$ and Mu$_{\text{BC}}^0$ and narrow the model down to a single candidate.
A representative source code and details of the curve fitting procedure are provided in the Supplemental Material~\cite{SM}.

\section{discussions}
We start by discussing data from the Pristine sample, where the single point defect, N$_\text{s}^0$, is expected to dominate the interaction with Mu.
The good-quality fitting, as shown in Fig.~\ref{fig:Fig_TS}(a) and (b), supports the model in Fig.~\ref{fig:Fig_Mu_diagrams}(a) as a good candidate for the Pristine sample.
This agreement can also be seen in the calculated repolarization curves as shown in Fig.~\ref{fig:Fig_int}(c), where the simulated time domain signal have been integrated and averaged.
Here, the data points for 290 and 20~K are broadly similar.
This observation is reflected in the obtained fit parameters, as summarized in TABLE~\ref{table:fit_results},
where we can draw a few physical insights.
First, prompt Mu yields, $f(\text{Mu}_\text{T}^0)$ $\approx$ 60~\% and $f(\text{Mu}_\text{BC}^0)$ $\approx$ 30~\%, were obtained at both temperatures.
This result agrees with the reported values~\cite{Connell}.
The negligibly small difference between 290 and 20~K may suggest that the Mu yields stay constant in this temperature range.
It is no wonder that these fractions stayed roughly the same in the NV sample because the process of NV generation cannot change bulk properties of the host crystal.

\begin{table*}
\centering
\caption{
Results of the global curve fitting.
}
\label{table:fit_results}
\begin{tabular}{c c c c c c c c c c}
\toprule
Sample & T [K] & $f(\text{Mu}_{\text{T}}^0)$ & $f(\text{Mu}_{\text{BC}}^0)$ & $\Lambda_{\text{T}}^0$ [MHz] & $\Lambda_{\text{BC}}^0$  [MHz] & $\Lambda_{\text{T/BC}}^{0/0}$  [MHz] & $\Lambda_{\text{BC/T}}^{0/0}$  [MHz] & $\Lambda_{\text{T/D}}^{0/-}$  [MHz] & $\Lambda_{\text{D/T}}^{-/0}$  [MHz] \\
\midrule
\multirow{2}{*}{Pristine}
& 290 & 0.600 $\pm$ 0.005  & 0.315 $\pm$ 0.007 & 58 $\pm$ 4 & 0.02 $\pm$ 0.02 & 1.19 $\pm$ 0.09 & 0.034  $\pm$ 0.002 & --- & ---\\
& 20 & 0.586 $\pm$ 0.004 & 0.324 $\pm$ 0.007 & 53 $\pm$ 3 & 0.22  $\pm$ 0.04 & 1.00 $\pm$ 0.06 & 0.034  $\pm$ 0.002 & --- & ---\\
\midrule
\multirow{2}{*}{NV}
& 290 & 0.632 $\pm$ 0.004 & 0.317 $\pm$ 0.005 & 73 $\pm$ 6 & 0.02 & 1.19 & 0.034 & 2.8 $\pm$ 0.4 & 0.065 $\pm$ 0.004\\
& 20 & 0.595 $\pm$ 0.004 & 0.355 $\pm$ 0.005 & 100 $\pm$ 12 & 0.22 & 1.00 & 0.034 & 9 $\pm$ 1 & 0.034 $\pm$ 0.003\\
\bottomrule
\end{tabular}
\end{table*}

Second, the values of $\Lambda_{T}^0$ are significantly large and interestingly comparable between the two temperatures.
This is the reason for the fast depolarization of Mu$_{\text{T}}^0$, as can be seen at the earlier times in Fig.~\ref{fig:Fig_TS}.
Since the dominant defect in the Pristine sample is N$_\text{s}^0$ centers, Mu$_{\text{T}}^0$ can exchange its electronic spin with the paramagnetic center upon scattering.
After the spin exchange and flipping, the muon spin in Mu$_{\text{T}}^0$ is depolarized via HF interaction.
Assuming that Mu$_{\text{T}}^0$ centers propagate in a wave-like manner, we consider $\Lambda_{T}^0$ as a collision rate with N$_\text{s}^0$ such that
\begin{equation}\label{eqn_transition}
\Lambda_{T}^0 = n v \sigma,
\end{equation}
where $n$ is the density of N$_\text{s}^0$ atoms,
$v$ is the effective velocity of Mu$_{\text{T}}^0$, and 
$\sigma$ is the scattering cross section between Mu$_{\text{T}}^0$ and N$_\text{s}^0$.
With the mobile Mu$_{\text{T}}^0$ and high-density N$_\text{s}^0$, Eq.~(\ref{eqn_transition}) describes how many times the Mu$_{\text{T}}^0 - \text{N}_\text{s}^0$ collision takes place per unit time.
\footnote{
This notation has been widely used to describe a Mu-electron (or hole) collision rate in semiconductors, $n_e v_e \sigma_e$,
where $n_e$ is the electronic density,
$v_e$ is the electron's thermal velocity, and
$\sigma_e$ is the scattering cross section~\cite{Kadono_Si_illumination, Yokoyama}.
Hence, $n_e v_e$ represents the electron's flux relative to Mu with its cross section defined by $\sigma_e$.
Here, we have adapted the notion to describe the Mu$_{\text{T}}^0 - \text{N}_\text{s}^0$ interaction.
}
Though $n$ can be estimated as $n = 2 \times 10^{19}$~$\text{cm}^{-3}$ in this sample, it is difficult to determine $v$ and $\sigma$ separately.
However, since N$_\text{s}^0$ is a deep donor state with its energy level located at 2.2~eV below the conduction band edge,
the unpaired electron should be tightly bound to the N$_\text{s}^0$ atom;
it is expected that its wavefunction, and hence $\sigma$, are nearly unchanged between 290 and 20~K.
As a result, $v$'s at the two temperatures are considered to be in the same magnitude.

Note that the diffusive motion of Mu$_{\text{T}}^0$ is governed by different physical processes in high and low temperatures~\cite{Storchak, Kadono_Rev}.
It is known that, above a crossover temperature, Mu$_{\text{T}}^0$ states are thermally activated and phonon-assisted tunneling is predominant for diffusion;
below the temperature, incoherent quantum tunneling becomes the dominant process.
For instance, the crossover temperature in GaAs is reportedly $\approx$100~K~\cite{Kadono_GaAs}.
The two diffusion processes can be equally effective, resulting in similar site hopping rates at low and high temperatures in a group of semiconductors and insulators.
In this context, $v$ at 290~K can be described with the equipartition theorem, whereas $v$ at 20~K should be described with the group velocity of Bloch wavefunctions.
The comparable $v$'s at 290 and 20~K can be plausible, but how does $v$ change (or not change) between the two temperatures?
We leave this question to a future temperature-dependent study.

In contrast to the mobile Mu$_{\text{T}}^0$ state, the slow rates on $\Lambda_{BC}^0$ agree with the immobile picture of Mu$_{\text{BC}}^0$,
although the value at 20~K is an order of magnitude larger than 290~K.
The Mu$_{\text{BC}}^0$ state can also hop between adjacent equivalent sites but with much slower rates than Mu$_{\text{T}}^0$.
This hopping behavior certainly depends on the temperature and can have the same driving mechanism as Mu$_{\text{T}}^0$ with a different crossover temperature.
Indeed, Mu$_{\text{BC}}^0$ in GaAs appears to have the crossover temperature around RT, resulting in a higher hopping rate in low temperatures~\cite{Yokoyama}.
For this reason, Mu$_{\text{BC}}^0$ in diamond presumably diffuses faster in low temperatures than RT.

Finally, the site exchange interaction between Mu$_{\text{T}}^0$ and Mu$_{\text{BC}}^0$,
whose rates are denoted by $\Lambda_{\text{T/BC}}^{0/0}$ and $\Lambda_{\text{BC/T}}^{0/0}$ with a slash between before and after a transition,
takes place commonly in tetrahedral semiconductors~\cite{Chow_Rev, Yokoyama, YokoyamaPRL}.
As shown in TABLE~\ref{table:fit_results}, the two orders of magnitude higher rate for $\Lambda_{\text{T/BC}}^{0/0}$ than $\Lambda_{\text{BC/T}}^{0/0}$ implies
the transition, Mu$_{\text{T}}^0$~$\rightarrow$~Mu$_{\text{BC}}^0$, to be predominant.
This observation agrees with the report by Odermatt et al.~\cite{Odermatt}, where they predicted this one-way transition and Mu$_{\text{BC}}^0$ to be the most stable site for all Mu states in diamond.
However, our analysis gave a small but finite rate for  $\Lambda_{\text{BC/T}}^{0/0}$, indicating the backward transition,
Mu$_{\text{BC}}^0$~$\rightarrow$~Mu$_{\text{T}}^0$ is still possible.

In contrast to the Pristine sample, $\mu$SR time spectra from the NV sample exhibited a remarkable difference in its response to the applied field, as shown in Fig.~\ref{fig:Fig_TS}(c) and (d);
this can be more clearly seen in the repolarization curves shown in Fig.~\ref{fig:Fig_int}(a) and (b) (also see Ref.~\cite{SM}).
The model in Fig.~\ref{fig:Fig_Mu_diagrams}(a) apparently fails to curve fit this data.
Since the system now has the NV centers as a site potentially interacting with Mu, it is natural to assume an additional path in the model.
Since the majority of NV centers in our system are in the negatively charged state, Mu states are expected to become diamagnetic after reacting with the NV$^-$ center.
In other words, the fast diffusing Mu$_{\text{T}}^0$ state can react with an electron-rich NV$^-$ center, presumably binding to it by forming a covalent bond.
This model is shown in Fig.~\ref{fig:Fig_Mu_diagrams}(b), where Mu$_{\text{D}}^-$ denotes the diamagnetic state with
the transition taking place in both directions {\it i.e.} Mu$_{\text{T}}^0$~$\rightleftarrows$~Mu$_{\text{D}}^-$.
By applying the aforementioned analysis procedure, this revised model can curve fit the data accurately.
This can be seen in the solid lines in Fig.~\ref{fig:Fig_TS}(c) and (d), and Fig.~\ref{fig:Fig_int}.
For the fitting, $\Lambda_{BC}^{0}$, $\Lambda_{T/BC}^{0/0}$, and $\Lambda_{BC/T}^{0/0}$ have been fixed to the parameters obtained from the Pristine sample to reduce the number of fit parameters.
This assumption is reasonable because they should be determined only by Mu interaction with the host crystal lattice.
The obtained parameters shown in TABLE~\ref{table:fit_results} gave consistent results,
with the prompt Mu yields and $\Lambda_{T}^{0}$ relaxation rates comparable with the Pristine sample.

The difference in the repolarzation behavior is, therefore, associated with the net transition converting Mu$_{T}^0$ into Mu$_{D}^-$.
Since $\Lambda_{T/D}^{0/-} \gg \Lambda_{D/T}^{-/0}$, the transition is essentially one-way, with the NV$^-$ centers acting as a trapping site for Mu$_{\text{T}}^0$.
Furthermore, the trapping seems to be more pronounced at 20~K, as shown in Fig.~\ref{fig:Fig_int}(d).
Indeed, TABLE~\ref{table:fit_results} shows a three-fold increase in $\Lambda_{T/D}^{0/-}$ with decreasing temperature.
To address physics behind this result, we consider the same relation as Eq.~(\ref{eqn_transition}) for $\Lambda_{T/D}^{0/-}$ {\it i.e.}
$\Lambda_{T/D}^{0/-} = n' v \sigma'$,
where $n'$ is the density of NV$^-$ centers, and 
$\sigma'$ is the cross section for a NV$^-$ center capturing Mu$_{\text{T}}^0$.
In principle, $n'$ is temperature dependent;
since NV centers are energetically more favorable in NV$^-$ than NV$^0$ states, a higher $n'$ is expected in a lower temperature.
However, in our sample, there is only a small room left for $n'$ to increase because of the nearly full conversion to NV$^-$ at RT.
Due to the aforementioned discussion, the effective Mu$_{\text{T}}^0$ velocity $v$ is considered to be comparable between 290 and 20~K.
Here, we note that there is a small but statistically significant increase in $\Lambda_{T}^{0}$ when compared with the Pristine data, which suggests an increase in $v$ in the NV sample for unknown reasons.
Nevertheless, the temperature dependence of $\Lambda_{T/D}^{0/-}$ may indicate that $\sigma'$ is temperature-dependent and larger at 20~K ---
this can be due to a wavefunction of the excess electron in NV$^-$ centers being delocalized due to quantum effects ({\it e.g.} tunneling and zero-point motion).
Further investigation, especially detailed temperature-dependent studies, would be necessary to give a clear answer.

\section{summary}
In this paper, we have studied the interaction of diffusive Mu$_{\text{T}}^0$ states with N$_\text{s}^0$ and NV$^-$ centers in type-Ib diamond.
Since the Mu interaction takes place predominantly with these point defects whose concentrations are roughly known, this system provides an ideal test bench to develop the Mu exchange model.
The numerical simulation with the density matrix method is key to deconvolute $\mu$SR time spectra into components and measure each Mu transition rate.
While there have been previous Mu studies on diamond, which measured Mu states, their yields, and interaction with the host crystal lattice, this paper identified the Mu state exchange dynamics in diamond for the first time.
For future development, a temperature-dependent study may be able to answer the questions about 1) the velocity of Mu$_{\text{T}}^0$, and 2) the Mu capture cross section of a NV$^-$ center.
Nevertheless, this study suggests its potential application to defect centers not only in diamond but also in other semiconducting systems.
There may be such applications in industrially important semiconductor systems, such as Si and SiC, where Mu$_{\text{T}}^0$ centers have been confirmed to exist in a stable form~\cite{Lichti}.

\section{acknowledgement}
This work was carried out using beamtime allocated by the STFC ISIS Facility~\cite{DataDoi}.
HA and TO acknowledge support by Quantum Leap Flagship Program (Q-LEAP; JPMXS 0118067395) of MEXT, Japan.
Finally, we are grateful for the assistance of a number of technical and support staff in the ISIS facility.


\end{document}